\documentclass[11pt]{article}
\usepackage[totalwidth=460truept,totalheight=600truept]{geometry}
\usepackage{latexsym,amssymb,amsmath,graphicx}
\usepackage[T1]{fontenc}
\usepackage{color}

\global\arraycolsep=1truept
\newcommand{\beq}{\begin{equation}}
\newcommand{\eeq}{\end{equation}}
\def\bea{\begin{eqnarray}}
\def\eea{\end{eqnarray}}
\definecolor{darkred}{rgb}{.8,0,0}

\definecolor{darkblu}{rgb}{0,0,.8}

\begin{document}



\vskip 1.4truecm

\begin{center}
{\huge \textbf{Magnetic Penrose Process and Blanford-Zanejk mechanism: A clarification}}

\vskip 1.5truecm

\textsl{Naresh Dadhich}

\textsl{Centre for Theoretical Physics, Jamia Millia Islamia, New Delhi 110 025}

\textit{\&}

\textsl{IUCAA, Post Bag 4, Ganeshkhind, Pune 411 007, India}

{\footnotesize nkd@iucaa.ernet.in}





\vskip 1truecm

\textbf{Abstract}
\end{center}

\bigskip

{\small The Penrose process (PP) is an ingenious mechanism of extracting rotational energy from a rotating black hole, however it was soon
realized that it was not very efficient for its astrophysical applications for powering the central engine of quasars and AGNs. The situation
however changed dramatically in the presence of magnetic field produced by the accretion disk surrounding the hole in the equatorial plane. In
1985, Wagh, Dhurandhar and Dadhich \cite{wdd} had for the first time considered the magnetic Penrose process (MPP) in which the magnetic field
could now provide the energy required for a fragment to ride on negative energy orbit thereby overcoming the stringent velocity constraint of
the original PP. Thus MPP turned very efficient and so much so that efficiency could now even exceed $100$ percent. They had in principle
established revival of PP for astrophysical applications in powering the high energy sources. MPP is however similar to the earlier discovered
and well known Blandford-Znajeck (BZ) mechanism in which the rotational energy of the hole is extracted out through a purely electromagnetic
process. Though both the processes use magnetic field as a device to extract rotational energy from the hole, yet their kernel is quite
different in spirit. For the former magnetic field provides the threshold energy for particle to get onto negative energy orbit so that the
other fragment goes out with enhanced energy while for the latter it generates an electric potential difference between the equatorial plane and
the polar region, and it is the discharge of which that drives the energy flux out of the hole. In other words, MPP is still rooted in the
spacetime geometry while BZ is essentially driven by electromagnetic interaction.}

\vskip 1truecm

\vfill\eject

\section{Introduction}

The Penrose process \cite{pen} is an ingenious mechanism of extracting rotational energy from a rotating black hole which was entirely driven
by the spacetime geometry. In the neighborhood of black hole event horizon, there occurs a region called ergosphere, which is bounded by the
static limit on one side and the horizon on the other, where there exist timelike particle orbits with negative total energy relative to infinity. Penrose exploited this
property for extraction of the hole's rotational energy. It was envisioned that a particle falling from infinity on to the hole splits into two
in the ergosphere, the one of which rides on the negative energy orbit and falls into the hole while the other escapes out with energy larger
than the incident particle. This is how the rotational energy of the black hole can be extracted out. This is what is known as the Penrose
process (PP). The essential requirement for it is existence of the ergosphere, where negative energy orbits can occur, and it is purely a
geometric construct caused by rotation of the hole. It cannot work for a non rotating black hole from which no energy could be extracted out by
this process simply because there can occur no negative energy orbits in its neighborhood. \\

It came at the very right time in 1969, just when the quasars, which were discovered a little earlier, and AGNs were asking for the energy
source for their central engine. Here was a process provided by the spacetime geometry itself and hence it created a good bit of excitement. It
was envisaged that a star like object grazing a super massive rotating black hole tidally or otherwise splits into fragments in the ergoshere,
some of which fall into the hole with negative energy while the others come out accelerated with large energy. However it was soon realized that
for a fragment to ride on the negative energy orbit, it should instantly be accelerated to velocity $>c/2$ \, \cite{bard, wald}. This was
however astrophysically unsustainable as there could be no conceivable process that could accelerate the fragments to such a high velocity
instantaneously. In other words, PP would have a very low efficiency of energy extraction and hence it is novel but not astrophysically viable.
Thus the excitement generated in 1969 died out by 1974. Since then some variants of it like the collisional PP were considered without much
improving the situation \cite{piran1, piran2}. \\

Then in 1977 came the celebrated Blandford-Zanejk (BZ) mechanism \cite{bz} of the electromagnetic extraction of the rotational energy of a
rotating black hole. Here it is envisioned that magnetic field produced by the accretion disk threads the black hole horizon and if the field is
strong enough, vacuum would be unstable to cascade production of electron-positron pairs and thereby establishing a force free magnetosphere.
Thus an electric potential difference would be generated between the equatorial plane and the polar region and when it is discharged, there
would be energy flux out of the hole. This is how the rotational energy of the hole would be extracted out by purely an electromagnetic process.
This was indeed one of the most promising astrophysically viable mechanisms for powering the central engine in quasars and AGNs. \\

In 1985 Wagh, Dhurandhar and Dadhich \cite{wdd} argued that a rotating black hole does not sit in isolation but is always surrounded by an
accretion disk which could produce a magnetic field threading the black hole horizon. It would therefore be pertinent to consider PP in presence
of electromagnetic field (magnetic field anchored on the inherent rotation in the ergosphere would also develop a quadrupole electric
component). Now the fragments in the ergosphere could be charged and they could therefore derive energy from electromagnetic field to get onto
the negative energy orbit thereby overcoming the stringent velocity constraint of the original PP. The process could thus be highly efficient
and hence it could be revived for astrophysical applications. This was what was exactly claimed in \cite{wdd}, and this was the first explicit
consideration of the magnetic Penrose process (MPP). It was first presented in \cite{iau, icgc} as electromagnetic revival of PP as energy
source for prime movers in high energy sources. The authors carried out a comprehensive study of the process establishing that efficiency could
even exceed $100$ percent in realistic settings \cite{pwdd, bdd, wagh}. It was concluded in writing a review on the energetics of black holes in
electromagnetic fields by the Penrose process \cite{review}. The viability of MPP was established in principle for a discrete particle accretion
and it had remained so until recently \cite{rn1, rn2}. This is indeed very gratifying to see that MPP is now considered to be the most pertinent
process for powering the central engine in quasars and AGNs. It is however thought that MPP is the same as BZ mechanism \cite{prague}, and that is
what I take up next.  \\

\section{MPP versus BZ: a synthesis}

It is generally believed that the two processes are similar and it would be hard to find a signature to distinguish between them. As mentioned
earlier, for both magnetic field plays the catalytic role for mining the rotational energy of the hole. MPP is rooted in the existence of
negative energy orbits which are caused by the coupling of particle angular momentum with the hole's angular momentum and in that magnetic field
adds its own contribution to effective angular momentum of the particle. It is that which facilitates particles of Newtonian orbits get onto
negative energy orbits. The magnetic field contribution thus favorably affects the particle energetics for MPP to work with very high
efficiency. On the other hand BZ is essentially an electromagnetic process with force free magnetosphere produced by cascade of
electron-positron pairs, and development of a potential difference between equatorial plane and polar region. Then the completion of electric
circuit by discharge of the potential difference that drives the (negative) energy flux into the hole and enhanced flux out to infinity. Here it
is the electromagnetic interaction which is at the core of this process. Since it takes place in the background geometry of a rotating black
hole which has inherent and irresistable rotation in the ergosphere, magnetic field lines are wound and anchored on the horizon. Spacetime
geometry thus plays a passive role in this case in contrast to MPP where it is an active driving element. This is how their kernel is quite
different in spirit but effectively both require very high magnetic field for the energy extraction to be efficient enough to serve as a viable
source for the high energy sources. Note that efficiency of MPP was shown to be greater than $100$ percent \cite{pwdd} which augurs very well
with the current investigations \cite{prague, rn1, rn2}. As a matter of fact efficiency exceeding $100$ percent could be considered as the
distinguishing feature of MPP.  \\

The distinction between the two could arise in the limiting case of magnetic field tending to zero. In that case MPP would go to PP, though
inefficient yet there is some possibility of energy extraction while for BZ there would be the null result. In the pertinent case of strong
magnetic field, one can perhaps take it as MPP or BZ, and it is simply a matter of taste whether one would prefer gravity/geometry or
electromagnetic field plays the active role. We could as well consider a synthesis of the two where one complements the other. For MPP, it is
envisaged that there is accreting matter from the disk on the hole that could provide fragments which could ride on negative energy orbits and
fall into the hole while others come out accelerated with large energy. The vacuum polarization of BZ could serve as a good viable source for
fragments in the ergosphere required for MPP to work. For BZ, the inherent irresistable rotation in the ergosphere plays a role in
enhancing the magnetic field strength and creating the magnetosphere. It would therefore be appropriate and pertinent to consider the synthesis
of the two mechanisms, and the energy extraction could be looked upon as MPP or BZ, the two equivalent processes for strong magnetic field
regime. We would just like to point out that MPP may be the same as BZ mechanism, it was however as such first considered by us in \cite{wdd} as revival of PP for astrophysical
applications. It is both heartening and gratifying to see that it is now the most favored mechanism for prime mover in the high energy sources
\cite{prague, rn1, rn2, ap}. \\

\section*{Acknowledgements} It is a plesaure to thank S Wagh, S Dhurandhar, S Parthasarthy and M Bhat, who formed a cohesive team that investigated various aspects of MPP, for successful collaboration. I would also like to thank several colleagues at various institutes and universities for discussions of this interesting phenomenon at the time of its discovery in mid to late 1980s. I am grateful to Ramesh Narayan for
referring to our work in his plenary lecture at the Relativity and Gravitation: 100 years after Einstein in Prague, June 25-29, 2012.

\end{document}